%% file: Main.tex
\documentclass[journal,12pt,onecolumn]{IEEEtran}

\usepackage{setspace}
\ifCLASSOPTIONonecolumn \doublespace \fi
\usepackage{graphics}
\usepackage{rotating}
\usepackage{amsfonts}
\usepackage{amsmath}
\usepackage{amsthm}
\usepackage{subfigure}
\usepackage{multirow}
\allowdisplaybreaks[1]
\newtheorem*{theorem}{Theorem}
\renewcommand{\IEEEQED}{\IEEEQEDopen}

\ifCLASSOPTIONonecolumn
\title{Bit-Interleaved Coded Multiple Beamforming with Perfect Coding}
\author{
Boyu Li and Ender Ayanoglu\\
Center for Pervasive Communications and Computing\\ 
Department of Electrical Engineering and Computer Science\\
The Henry Samueli School of Engineering\\
University of California, Irvine\\
Irvine, California 92697-2625\\
Email: boyul@uci.edu, ayanoglu@uci.edu}
\date{} 
\else
\title{Bit-Interleaved Coded Multiple Beamforming with Perfect Coding}
\author{
Boyu Li and Ender Ayanoglu\\
\\
Center for Pervasive Communications and Computing\\ 
Department of Electrical Engineering and Computer Science\\
The Henry Samueli School of Engineering\\
University of California, Irvine\\
Irvine, California 92697-2625\\
Email: boyul@uci.edu, ayanoglu@uci.edu}
\date{} 
\fi

\begin{document}
\maketitle

\ifCLASSOPTIONonecolumn
 \setlength\arraycolsep{4pt}
\else
 \setlength\arraycolsep{2pt}
\fi

\input{Abstract}
\input{Introduction}
\input{System_model}
\input{Analysis}
\input{Decoding}
\input{Results}
\input{Conclusion}
\bibliographystyle{IEEEtran}
\bibliography{IEEEabrv,Mybib}
\end{document}

%% file: Abstract.tex
\begin{abstract}

When the Channel State Information (CSI) is known by both the transmitter and the receiver, beamforming techniques employing Singular Value Decomposition (SVD) are commonly used in Multiple-Input Multiple-Output (MIMO) systems. Without channel coding, there is a trade-off between full diversity and full multiplexing. When channel coding is added, both of them can be achieved as long as the code rate $R_c$ and the number of employed subchannels $S$ satisfy the condition $R_cS \leq 1$. By adding a properly designed constellation precoder, both  full diversity and full multiplexing can be achieved for both uncoded and coded systems with the trade-off of a higher decoding complexity, e.g., Fully Precoded Multiple Beamforming (FPMB) and Bit-Interleaved Coded Multiple Beamforming with Full Precoding (BICMB-FP) without the condition $R_cS \leq 1$. Recently discovered Perfect Space-Time Block Code (PSTBC) is a full-rate full-diversity space-time code, which achieves efficient shaping and high coding gain for MIMO systems. In this paper, a new technique, Bit-Interleaved Coded Multiple Beamforming with Perfect Coding (BICMB-PC), is introduced. BICMB-PC transmits PSTBCs through convolutional coded SVD systems. Similarly to BICMB-FP, BICMB-PC achieves both full diversity and full multiplexing, and its performance is almost the same as BICMB-FP. The advantage of BICMB-PC is that it can provide a much lower decoding complexity than BICMB-FP, since the real and imaginary parts of the received signal can be separated for BICMB-PC of dimensions $2$ and $4$, and only the part corresponding to the coded bit is required to acquire one bit metric for the Viterbi decoder.

\end{abstract}

%% file: Introduction.tex
\section{Introduction} \label{sec:Introduction}


When Channel State Information (CSI) is available at both the transmitter and the receiver, beamforming techniques exploiting Singular Value Decomposition (SVD) are applied in a Multiple-Input Multiple-Output (MIMO) 
system to achieve spatial multiplexing\footnotemark \footnotetext{In this paper, the term ``spatial multiplexing" is used to describe the number of spatial subchannels, as in \cite{Paulraj_ST}. Note that the term is different from ``spatial multiplexing gain" defined in \cite{Zheng_DM}.} and thereby increase the data rate, or to enhance the performance \cite{Jafarkhani_STC}. However, spatial multiplexing without channel coding results in the loss of the full diversity order \cite{Sengul_DA_SMB}. To overcome the diversity degradation, Bit-Interleaved Coded Multiple Beamforming (BICMB), which interleaves the codewords through the multiple subchannels with different diversity orders, was proposed \cite{Akay_BICMB}, \cite{Akay_On_BICMB}. BICMB can achieve both full diversity and full multiplexing as long as the code rate $R_c$ and the number of employed subchannels $S$ satisfy the condition $R_cS \leq 1$ \cite{Park_DA_BICMB}, \cite{Park_DA_BICMB_J}. In \cite{Mohammed_DIA}, \cite{Mohammed_XY}, X-Codes and Y-Codes were introduced to increase the diversity of multiple beamforming which transmits multiple streams. These techniques do not guarantee full diversity when the number of transmit or receive antennas is larger than $2$, and require relatively high precoding complexity. In \cite{Park_CPB}, \cite{Park_BICMB_CP}, \cite{Park_MB_CP}, \cite{Park_CPMB}, it was shown that by employing the constellation precoding technique, which has very low precoding complexity, full diversity and full multiplexing can be achieved simultaneously for both uncoded and convolutional coded SVD systems with the trade-off of a higher decoding complexity. Specifically, in the uncoded case, full diversity requires that all streams are precoded, i.e., Fully Precoded Multiple Beamforming (FPMB). A similar result was reported in \cite{Srinivas_COISM_CSI} with a technique employing the rotated Quadrature Amplitude Modulation (QAM) constellation. On the other hand, for the convolutional coded SVD systems without the condition $R_cS \leq 1$, other than full precoding, i.e., Bit-Interleaved Coded Multiple Beamforming with Full Precoding (BICMB-FP), partial precoding, i.e., Bit-Interleaved Coded Multiple Beamforming with Partial Precoding (BICMB-PP) could also achieve both full diversity and full multiplexing with the properly designed combination of the convolutional code, the bit interleaver, and the constellation precoder.

In \cite{Oggier_PSTBC}, the Perfect Space-Time Block Code (PSTBC) was introduced for dimensions $2$, $3$, $4$, and $6$. PSTBCs have the full rate, full diversity, nonvanishing minimum determinant for increasing spectral efficiency, uniform average transmitted energy per antenna, good shaping of the constellation, and high coding gain. In \cite{Elia_PSTBC}, PSTBCs were generalized to any dimension. However, it was proved in \cite{Berhuy_PSTC} that particular PSTBCs, yielding increased coding gain, only exist in dimensions $2$, $3$, $4$, and $6$. Due to the advantages of PSTBCs, the Golden Code (GC), which is the best known PSTBC for MIMO systems with two transmit and two receive antennas \cite{Belfiore_GC}, \cite{Dayal_STC}, has been incorporated into the $802.16$e Worldwide Interoperability for Microwave Access (WiMAX) standard \cite{IEEE_802_16e}.

In our previous work \cite{Li_GCMB}, Perfect Coded Multiple Beamforming (PCMB) was proposed. PCMB combines PSTBCs with uncoded multiple beamforming, and achieves full diversity, full multiplexing, and full rate at the same time, in a similar fashion to a MIMO system employing PSTBC and FPMB. It was shown that for dimensions $2$ and $4$, all these three techniques have close Bit Error Rate (BER) performance, while the worst-case decoding complexity of PCMB is significantly less than a MIMO system employing PSTBC for both low and high Signal-to-Noise Ratio (SNR), and is much lower than FPMB for low SNR, which provides the advantage of PCMB. 

In this paper, a new technique with both full diversity and full multiplexing for convolutional coded SVD systems, Bit-Interleaved Coded Multiple Beamforming with Perfect Coding (BICMB-PC), is proposed. BICMB-PC transmits bit-interleaved codewords of PSTBC, instead of PSTBC codewords without channel coding for PCMB, through the multiple subchannels. Diversity analysis of BICMB-PC is carried out to prove that it achieves the full diversity order. Simulation results show that BICMB-PC achieves almost the same BER performance as BICMB-FP, which is also a technique for convolutional coded SVD systems with both full diversity and full multiplexing. Moreover, the decoding complexity analysis shows the advantage of BICMB-PC, which has much lower complexity than BICMB-FP for both low and high SNR in dimensions $2$ and $4$. The reason is that the real and imaginary parts of the received signal of BICMB-PC can be separated, which is not applied for BICMB-FP, and only the part corresponding to the coded bit is required to calculate one bit metric for the Viterbi decoder. Compared to the uncoded system, the complexity reduction from BICMB-FP to BICMB-PC is greater than the reduction from FPMB to PCMB achieved in \cite{Li_GCMB} for dimensions $2$ and $4$. Moreover, since the precoded part of BICMB-PP could be considered as a smaller dimensional BICMB-FP, BICMB-PC of dimensions $2$ and $4$ could be applied to replace the precoded part and reduce the complexity for BICMB-PP. 

The remainder of this paper is organized as follows: In Section \ref{sec:System_model}, the description of BICMB-PC is given. In Section \ref{sec:Analysis}, the diversity analysis of BICMB-PC is provided. In Section \ref{sec:Decoding}, the decoding technique and complexity analysis of BICMB-PC are shown. In Section \ref{sec:Results}, simulation results are shown. Finally, a conclusion is provided in Section \ref{sec:Conclusion}.

%% file: System_model.tex
\section{BICMB-PC Overview} \label{sec:System_model}

\ifCLASSOPTIONonecolumn
\begin{figure}[!m]
\centering \includegraphics[width = 1.0\linewidth]{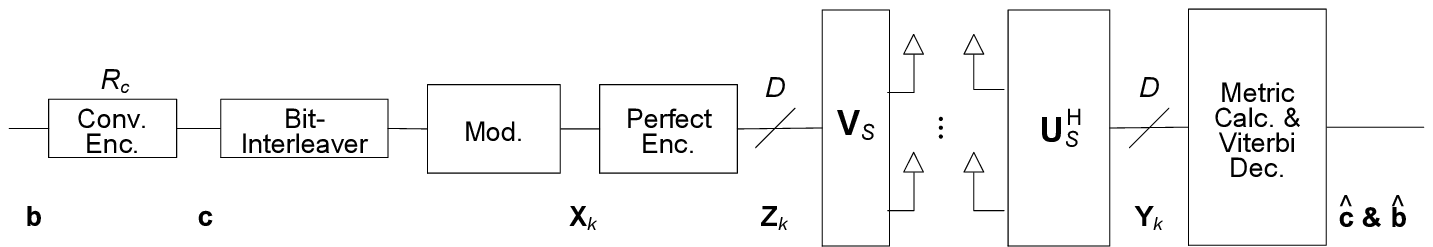}
\caption{Structure of BICMB-PC.} \label{fig:system_model}
\end{figure}
\else
\begin{figure}[!t]
\centering \includegraphics[width = 1.0\linewidth]{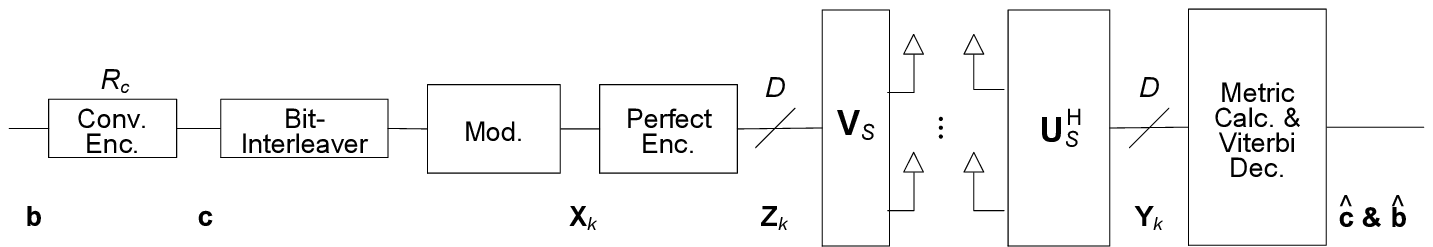}
\caption{Structure of BICMB-PC.} \label{fig:system_model}
\end{figure}
\fi

The structure of BICMB-PC is presented in Fig. \ref{fig:system_model}. First, the convolutional encoder of code rate $R_c$, possibly combined with a perforation matrix \cite{Haccoun_PCC} for a high rate punctured code, generates the bit codeword $\mathbf{c}$ from the information bits. Then, a random bit-interleaver is applied to generate the interleaved bit sequence, which is then modulated by $M$-QAM or $M$-HEX \cite{Forney_EM} and mapped by Gray encoding. Then $D^2$ consecutive complex-valued scalar symbols are encoded into one PSTBC codeword, where $D \in \{2,3,4,6 \}$ is the system dimension. Hence, the $k$th PSTBC codeword $\mathbf{Z}_k$ is constructed as
\begin{equation}
\mathbf{Z}_k=\sum_{v=1}^D{\mathrm{diag}(\mathbf{G}\mathbf{x}_{v,k})\mathbf{E}^{v-1}},
\label{eq:PSTBC}
\end{equation}
where $\mathbf{G}$ is an $D\times D$ unitary matrix, $\mathbf{x}_{v,k}$ is an $D\times1$ vector whose elements are the $v$th $D$ input modulated scalar symbols, and 
\begin{align*}
\mathbf{E} = \left[
\begin{array}{ccccc}
0 & 1 & \cdots & 0 & 0 \\
0 & 0 & 1 & \cdots & \vdots \\
\vdots & \vdots & \ddots & \ddots & \vdots \\
0 & \cdots & \cdots & \cdots & 1 \\
g & 0 & \cdots & 0 & 0
\end{array}
\right],
\end{align*}
with 
\begin{align*}
g = \left\lbrace 
\begin{array}{cc}
i, & D=2,4, \\
e^{\frac{2{\pi}i}{3}}, & D=3, \\
-e^{\frac{2{\pi}i}{3}}, & D=6,
\end{array} \right.
\end{align*}
and $\mathrm{diag}(\mathbf{w}=[w_1, \ldots, w_D]^T)$ denotes a diagonal matrix with diagonal entries $w_1, \ldots, w_D$. The selection of the $\mathbf{G}$ matrix for different dimensions can be found in \cite{Oggier_PSTBC}. 

The MIMO channel $\mathbf{H} \in \mathbb{C}^{N_r \times N_t}$ is assumed to be quasi-static, Rayleigh, and flat fading, and known by both the transmitter and the receiver, where $N_r$ and $N_t$ denote the number of receive and transmit antennas respectively, and $\mathbb{C}$ stands for the set of complex numbers. The beamforming matrices are determined by the SVD of the MIMO channel, i.e., $\mathbf{H} = \mathbf{U \Lambda V}^H$, where $\mathbf{U}$ and $\mathbf{V}$ are unitary matrices, and $\mathbf{\Lambda}$ is a diagonal matrix whose $s$th diagonal element, $\lambda_s \in \mathbb{R}^+$, is a singular value of $\mathbf{H}$ in decreasing order, where $\mathbb{R}^+$ denotes the set of positive real numbers. When $S$ streams are transmitted at the same time, the first $S$ vectors of $\mathbf{U}$ and $\mathbf{V}$ are chosen to be used as beamforming matrices at the receiver and the transmitter, respectively. In the case of BICMB-PC, $N_r=N_t=S=D$.

The received signal corresponding to the $k$th PSTBC codeword is
\begin{align}
\mathbf{Y}_{k} = \mathbf{{\Lambda}} \mathbf{Z}_k+\mathbf{N}_k, \label{eq:detected_matrix}
\end{align}
where $\mathbf{Y}_k$ is an $D \times D$ complex-valued matrix, and $\mathbf{N}_k$ is the $D \times D$ complex-valued additive white Gaussian noise matrix whose elements have zero mean and variance $N_0 = D / SNR$. The channel matrix $\mathbf{H}$ is complex Gaussian with zero mean and unit variance. The total transmitted power is scaled as $D$ in order to make the received SNR $SNR$.

The location of the coded bit $c_{k'}$ within the PSTBC codeword sequence is denoted as $k' \rightarrow (k, (m,n), j)$, where $k$, $(m,n)$, and $j$ are the index of the PSTBC codewords, the symbol position in $\mathbf{X}_k=[\mathbf{x}_{1,k}, \ldots, \mathbf{x}_{D,k}]$, and the bit position on the label of the scalar symbol $x_{(m,n),k}$, respectively. Let $\chi$ denote the signal set of the modulation scheme, and let $\chi_b^j$ denote a subset of $\chi$ whose labels have $b \in \{0, 1\}$ in the $j$th bit position. Define the one-to-one mapping from $\mathbf{X}_k$ to $\mathbf{Z}_k$ as $\mathbf{Z}_k = \mathbb{M} \{ \mathbf{X}_k \}$. By using the location information and the input-output relation in (\ref{eq:detected_matrix}), the receiver calculates the Maximum Likelihood (ML) bit metrics for $c_{k'}=b$ as
\begin{align}
\Gamma^{(m,n),j}(\mathbf{Y}_k, c_{k'}) = \min_{\mathbf{X} \in \eta_{c_{k'}}^{(m,n),j}} \| \mathbf{Y}_k - \mathbf{\Lambda} \mathbb{M} \{ \mathbf{X} \} \|^2, \label{eq:ML_bit_metrics}
\end{align}
where $\eta_{c_{k'}}^{(m,n),j}$ is defined as 
\begin{align*}
\eta_{b}^{(m,n),j} = \{ \mathbf{X}: x_{(u,v)=(m,n)} \in \chi_{b}^{j}, \, \mathrm{and} \, x_{(u,v) \neq (m.n)} \in \chi \}.
\end{align*}

Finally, the ML decoder, which uses the soft-input Viterbi decoding \cite{Lin_ECC} to find a codeword with the minimum sum weight, makes decisions according to the rule given by \cite{Caire_BICM} as
\begin{align}
\mathbf{\hat{c}} = \arg\min_{\mathbf{c}} \sum_{k'} \Gamma^{(m,n),j}(\mathbf{Y}_k, c_{k'}).
\label{eq:Decision_Rule}
\end{align}

%% file: Analysis.tex
\section{Diversity Analysis} \label{sec:Analysis}


Based on the bit metrics in (\ref{eq:ML_bit_metrics}), the instantaneous Pairwise Error Probability (PEP) between the transmitted codeword $\mathbf{c}$ and the decoded codeword $\mathbf{\hat{c}}$ is
\begin{align}
\mathrm{Pr} \left( \mathbf{c} \rightarrow \hat{\mathbf{c}} \mid \mathbf{H} \right) = \mathrm{Pr} \left( \sum_{k'} \min_{\mathbf{X} \in \eta_{c_{k'}}^{(m,n),j}} \| \mathbf{Y}_k - \mathbf{\Lambda} \mathbb{M} \{ \mathbf{X} \} \|^2 \geq \right. \ifCLASSOPTIONtwocolumn \nonumber \\ \fi \left. \sum_{k'} \min_{\mathbf{X} \in \eta_{\hat{c}_{k'}}^{(m,n),j}} \| \mathbf{Y}_k - \mathbf{\Lambda} \mathbb{M} \{ \mathbf{X} \} \|^2 \mid \mathbf{H} \right). \label{eq:PEP_original}
\end{align}
Let $d_H$ denote the Hamming distance between $\mathbf{c}$ and $\hat{\mathbf{c}}$. 
Since the bit metrics corresponding to the same coded bits between the pairwise errors are the same, (\ref{eq:PEP_original}) is rewritten as
\begin{align}
\mathrm{Pr} \left(\mathbf{c} \rightarrow \hat{\mathbf{c}} \mid \mathbf{H}\right) = \mathrm{Pr} \left( \sum_{k', d_H} \min_{\mathbf{X} \in \eta_{c_{k'}}^{(m,n),j}} \| \mathbf{Y}_k - \mathbf{\Lambda} \mathbb{M} \{ \mathbf{X} \} \|^2 \geq \right. \ifCLASSOPTIONtwocolumn \nonumber \\ \fi \left. \sum_{k', d_H} \min_{\mathbf{X} \in \eta_{\hat{c}_{k'}}^{(m,n),j}} \| \mathbf{Y}_k - \mathbf{\Lambda} \mathbb{M} \{ \mathbf{X} \} \|^2 \mid \mathbf{H} \right),
\label{eq:PEP_for_different_codedbits}
\end{align}
where $\sum_{k', d_H}$ stands for the summation of the $d_H$ values corresponding to the different coded bits between the bit codewords.

Define $\tilde{\mathbf{X}}_k$ and $\hat{\mathbf{X}}_k$ as
\begin{equation}
\begin{split}
\tilde{\mathbf{X}}_k = \arg \min_{\mathbf{X} \in \eta_{c_{k'}}^{(m,n),j}} \| \mathbf{Y}_k - \mathbf{\Lambda} \mathbb{M} \{ \mathbf{X} \} \|^2, \\ 
\hat{\mathbf{X}}_k = \arg \min_{\mathbf{X} \in \eta_{\bar{c}_{k'}}^{(m,n),j}} \| \mathbf{Y}_k - \mathbf{\Lambda} \mathbb{M} \{ \mathbf{X} \} \|^2,
\end{split}
\label{eq:arg_min}
\end{equation}
where $\bar{c}_{k'}$ is the complement of $c_{k'}$ in binary. It is easily found that $\tilde{\mathbf{X}}_k$ is different from $\hat{\mathbf{X}}_k$ since the sets that $x_{(m,n)}$ belong to are disjoint, as can be seen from the definition of $\eta_{c_{k'}}^{(m,n),j}$. In the same manner, it is clear that $\mathbf{X}_k$ is different from $\hat{\mathbf{X}}_k$. With $\tilde{\mathbf{Z}}_k = \mathbb{M} \{ \tilde{\mathbf{X}}_k \}$ and $\hat{\mathbf{Z}}_k=\mathbb{M} \{ \hat{\mathbf{X}}_k \}$, (\ref{eq:PEP_for_different_codedbits}) is rewritten as
\begin{align}
\mathrm{Pr} \left( \mathbf{c} \rightarrow \hat{\mathbf{c}} \mid \mathbf{H} \right) = \mathrm{Pr} \left( \sum_{k', d_H} \| \mathbf{Y}_k - \mathbf{\Lambda} \tilde{\mathbf{Z}}_k \|^2 \geq \right. \ifCLASSOPTIONtwocolumn \nonumber \\ \fi \left. \sum_{k', d_H} \| \mathbf{Y}_k - \mathbf{\Lambda} \hat{\mathbf{Z}}_k \|^2 \right).
\label{eq:alt_expression_PEP_diffbits}
\end{align}
Based on the fact that $\| \mathbf{Y}_k - \mathbf{\Lambda} \mathbf{Z}_k \|^2 \geq  \| \mathbf{Y}_k - \mathbf{\Lambda} \tilde{\mathbf{Z}}_k \|^2$ and the relation in (\ref{eq:detected_matrix}), equation (\ref{eq:alt_expression_PEP_diffbits}) is upper-bounded by
\begin{align}
\mathrm{Pr} (\mathbf{c} \rightarrow \hat{\mathbf{c}} \mid \mathbf{H}) \leq \mathrm{Pr} \left( \xi \geq \sum_{k', d_H} \| \mathbf{\Lambda} (\mathbf{Z}_k - \hat{\mathbf{Z}}_k) \|^2 \right), \label{eq:PEP_upperbounded}
\end{align}
where $\xi = \sum_{k', d_H} \mathrm{Tr} [ - (\mathbf{Z}_k - \hat{\mathbf{Z}_k})^H \mathbf{\Lambda}^H \mathbf{N}_k- \mathbf{N}_k^H \mathbf{\Lambda} (\mathbf{Z}_k - \hat{\mathbf{Z}}_k) ]$. Since $\xi$ is a zero-mean Gaussian random variable with variance $2 N_0 \sum_{k', d_H} \| \mathbf{\Lambda}  (\mathbf{Z}_k - \hat{\mathbf{Z}}_k) \| ^2$, (\ref{eq:PEP_upperbounded}) is replaced by the $Q$ function as
\begin{align}
\mathrm{Pr} (\mathbf{c} \rightarrow \hat{\mathbf{c}} \mid \mathbf{H}) \leq \mathrm{Q} \left( \sqrt \frac{\sum_{k', d_H} \| \mathbf{\Lambda} (\mathbf{Z}_k - \hat{\mathbf{Z}}_k) \|^2}{2N_0}
\right). \label{eq:PEP_Q}
\end{align}
By using the upper bound on the $Q$ function $Q(x) \leq \frac{1}{2} e^{-x^2/2}$, the average PEP can be upper bounded as
\begin{align}
\mathrm{Pr} \left( \mathbf{c} \rightarrow \hat{\mathbf{c}} \right) &= E \left[ \mathrm{Pr} \left( \mathbf{c} \rightarrow \hat{\mathbf{c}} \mid \mathbf{H} \right) \right] \nonumber \\ 
&\leq E \left[ \frac{1}{2} \exp \left(- \frac{\sum_{k', d_H} \| \mathbf{\Lambda} (\mathbf{Z}_k - \hat{\mathbf{Z}}_k) \| ^2}{4 N_0}
\right) \right]. \label{eq:PEP_average}
\end{align}

In \cite{Li_GCMB}, it was shown that
\begin{align}
\| \mathbf{\Lambda} \mathbf{Z}_k \|^2 &=  \mathrm{Tr} [ \mathbf{Z}_k^H \mathbf{\Lambda}^H \mathbf{\Lambda} \mathbf{Z}_k ] \nonumber \\ 
&= \sum_{u=1}^{S} {\lambda}_u^2 \sum_{v=1}^{D} | \mathbf{g}^T_u \mathbf{x}_{v,k} |^2,
\label{eq:Lambda_Z_square} 
\end{align}
where $\mathbf{g}^T_u$ denotes the $u$th row of $\mathbf{G}$. By replacing $\mathbf{Z}_k$ in (\ref{eq:Lambda_Z_square}) by $\mathbf{Z}_k-\hat{\mathbf{Z}}_k$, (\ref{eq:PEP_average}) is then rewritten as
\begin{align}
\mathrm{Pr} \left( \mathbf{c} \rightarrow \hat{\mathbf{c}} \right) &\leq E \left[ \frac{1}{2} \exp \left(- \frac{\sum_{k', d_H} \sum_{u=1}^{D} \lambda_u^2 \rho_{u,k}} {4 N_0} \right) \right] \nonumber \\
 & = E \left[ \frac{1}{2} \exp \left(- \frac{ \sum_{u=1}^{D} \lambda_u^2 \sum_{k', d_H} \rho_{u,k}} {4 N_0} \right) \right], 
\label{eq:PEP_average_2}
\end{align}
where 
\begin{align}
\rho_{u,k}=\sum_{v=1}^{D} | \mathbf{g}^T_u (\mathbf{x}_{v,k} - \hat{\mathbf{x}}_{v,k})|^2.
\label{eq:tau}
\end{align} 
The upper bound in (\ref{eq:PEP_average_2}) can be further bounded by employing a theorem from \cite{Park_UP_MPDF} which is given below.

\begin{theorem}
Consider the largest $S \leq \min(N_t, N_r)$ eigenvalues $\mu_s$ of the uncorrelated central $N_r \times N_t$ Wishart matrix that are sorted in decreasing order, and a weight vector $\boldsymbol{\rho} = [\rho_1, \cdots, \rho_S]^T$ with
non-negative real elements. In the high SNR regime, an upper bound for the expression $E [ \exp (-\gamma
\sum_{s=1}^S \rho_s \mu_s ) ]$, which is used in the diversity analysis of a number of MIMO systems, is
\begin{align*}
E\left[ \exp \left( - \gamma \sum\limits_{s=1}^S \rho_s \mu_s \right) \right] \leq \zeta \left( \rho_{min} \gamma \right)^{-(N_r-\delta+1)(N_t-\delta+1)},
\end{align*}
where $\gamma$ is SNR, $\zeta$ is a constant, $\rho_{min} = \min_{\rho_i \neq 0} {\{ \rho_i \}}_{i=1}^{S}$, and $\delta$
is the index to the first non-zero element in the weight vector.
\label{theorem:E_PEP}
\end{theorem}
\begin{IEEEproof}
See \cite{Park_UP_MPDF}.
\end{IEEEproof}

Based on the aforementioned theorem, full diversity is achieved if and only if $\delta = 1$, which is equivalent to $\rho_1 > 0$. Note that $\rho_{1,k} > 0$ in (\ref{eq:tau}) because all elements in $\mathbf{g}^T_1$ are nonzero \cite{Oggier_PSTBC}, and therefore $\delta = 1$. By applying the Theorem to (\ref{eq:PEP_average_2}), an upper bound of PEP is
\begin{align}
\mathrm{Pr} \left( \mathbf{c} \rightarrow \hat{\mathbf{c}} \right) &\leq \zeta \left( \frac{\min\{\sum_{k', d_H} \rho_{u,k}\}}{4 D} SNR \right)^{-N_rN_t}.
\label{eq:PEP_PSB_final}
\end{align}
Hence, BICMB-PC achieves the full diversity order.

%% file: Decoding.tex
\section{Decoding} \label{sec:Decoding}

It was shown in \cite{Li_GCMB} that each element of $\mathbf{\Lambda} \mathbf{Z}_k$ in (\ref{eq:detected_matrix}) is related to only one of the $\mathbf{x}_{v,k}$. Consequently, the elements of $\mathbf{\Lambda} \mathbf{Z}_k$ can be divided into $D$ groups, where the $v$th group contains elements related to $\mathbf{x}_{v,k}$, and $v=1, \cdots, D$. 

Take GC ($D=2$) as an example,
\begin{align}
\mathbf{\Lambda} \mathbf{Z}_k =  
\left[ \begin{array}{cc} 
{\lambda}_1\mathbf{g}^T_1\mathbf{x}_{1,k} & {\lambda}_1\mathbf{g}^T_1\mathbf{x}_{2,k} \\
i{\lambda}_2\mathbf{g}^T_2\mathbf{x}_{2,k} & {\lambda}_2\mathbf{g}^T_2\mathbf{x}_{1,k}
\end{array} \right].
\label{eq:Lambda_Z} 
\end{align}
The input-output relation in (\ref{eq:detected_matrix}) is then decomposed into two equations as
\begin{align}
\begin{split}
&\mathbf{\breve{y}}_{1,k}= \left[
\begin{array}{c} 
Y_{(1,1),k} \\
Y_{(2,2),k}
\end{array} \right]  = \left[
\begin{array}{c}
{\lambda}_1\mathbf{g}^T_1\mathbf{x}_{1,k} \\
{\lambda}_2\mathbf{g}^T_2\mathbf{x}_{1,k}
\end{array} \right] + \left[
\begin{array}{c} 
N_{(1,1),k} \\
N_{(2,2),k}
\end{array} \right], \\
&\mathbf{\breve{y}}_{2,k} = \left[
\begin{array}{c} 
Y_{(1,2),k} \\
Y_{(2,1),k}
\end{array} \right]  = \left[
\begin{array}{c}
{\lambda}_1\mathbf{g}^T_1\mathbf{x}_{2,k} \\
i{\lambda}_2\mathbf{g}^T_2\mathbf{x}_{2,k}
\end{array} \right] + \left[
\begin{array}{c} 
N_{(1,2),k} \\
N_{(2,1),k}
\end{array} \right],
\end{split} \label{eq:deteced_symbol_decomposed}
\end{align} 
where $Y_{(m,n),k}$ and $N_{(m,n),k}$ denote the $(m,n)$th element of $\mathbf{Y}_k$ and $\mathbf{N}_k$ respectively. Let $\mathbf{\breve{n}}_{1,k}=[N_{(1,1),k}, N_{(2,2),k}]^T$ and $\mathbf{\breve{n}}_{2,k}=[N_{(1,2),k}, N_{(2,1),k}]^T$, then (\ref{eq:deteced_symbol_decomposed}) can be further rewritten as
\begin{align}
\begin{split}
&\breve{\mathbf{y}}_{1,k} = \mathbf{\Lambda G} \mathbf{x}_{1,k} + \breve{\mathbf{n}}_{1,k}, \\
&\breve{\mathbf{y}}_{2,k} = \mathbf{\Phi \Lambda G} \mathbf{x}_{2,k} + \breve{\mathbf{n}}_{2,k},
\end{split} \label{eq:deteced_symbol_decomposed_2}
\end{align} 
where 
\begin{align*}
\mathbf{\Phi} = \left[\
\begin{array}{cc}
1 & 0 \\
0 & i
\end{array} \right].
\end{align*}

A similar procedure can be applied to larger dimensions. Then in general, the received signal, which is divided into $D$ parts, can be represented as
\begin{align}
\breve{\mathbf{y}}_{v,k} = \mathbf{\Phi}_v \mathbf{\Lambda G} \mathbf{x}_{v,k} + \breve{\mathbf{n}}_{v,k}, 
\label{eq:deteced_symbol_decomposed_3}
\end{align} 
where  $v=1, \ldots, D$ and $\mathbf{\Phi}_v=\mathrm{diag}(\phi_{v,1}, \ldots, \phi_{v,D})$ is a diagonal unitary matrix whose elements satisfy
\begin{align*}
{\phi}_{v,u} = \left\lbrace 
\begin{array}{cc}
1, & 1 \leq u \leq D+1-v, \\
g, & D+2-v \leq u \leq D.
\end{array} \right.
\end{align*} 

By using the QR decomposition of $\mathbf{\Lambda G}=\mathbf{Q} \mathbf{R}$, where $\mathbf{R}$ is an upper triangular matrix, and the matrix $\mathbf{Q}$ is unitary, and moving $\mathbf{\Phi}_v\mathbf{Q}$ to the left hand, (\ref{eq:deteced_symbol_decomposed_3}) is rewritten as 
\begin{align}
\tilde{\mathbf{y}}_{v,k} = \mathbf{Q}^H \mathbf{\Phi}_v^H \breve{\mathbf{y}}_{v,k} = \mathbf{R}\mathbf{x}_{v,k} + \tilde{\mathbf{n}}_{v,k},
\label{eq:deteced_symbol_decomposed_4}
\end{align}
where $\tilde{\mathbf{n}}_{v,k} = \mathbf{Q}^H \mathbf{\Phi}_v^H \breve{\mathbf{n}}_{v,k}$. Then the ML bit metrics in (\ref{eq:ML_bit_metrics}) can be simplified as
\begin{align}
\Gamma^{(m,n),j}(\mathbf{Y}_k, c_{k'}) = \min_{\mathbf{x} \in \xi_{c_{k'}}^{n,j}} \| \tilde{\mathbf{y}}_{m,k} - \mathbf{R} \mathbf{x} \|^2, \label{eq:ML_bit_metrics_2}
\end{align}
where $\xi_{c_{k'}}^{n,j}$ is a subset of $\chi^D$, defined as
\begin{align*}
\xi_{b}^{n,j} = \{ \mathbf{x} = [x_1 \, \cdots \, x_D ]^T : x_{d=n} \in \chi_{b}^{j}, \, \mathrm{and} \, x_{d \neq n} \in \chi \}.
\end{align*}

The simplified ML bit metrics (\ref{eq:ML_bit_metrics_2}) are similar to BICMB-FP presented in \cite{Park_BICMB_CP}, \cite{Park_MB_CP}, \cite{Park_CPMB}, which are used to calculate ${1\over 2}M^D$ points by exhaustive search for one bit metric. Hence, the complexity is proportional to $M^{D}$, denoted by $\mathcal{O}(M^{D})$. Sphere Decoding (SD) is an alternative for ML with reduced complexity \cite{Jalden_SD}, which reduces the average complexity and provides the worst-case complexity of $\mathcal{O}(M^{D})$. Moreover, if an efficient implementation of a slicer \cite{Sinnokrot_STBC_LMLDC} is applied, the worst-case complexity is then $\mathcal{O}(M^{D}-1)$.

Particularly, it was proved in \cite{Li_GCMB} that $\mathbf{R}$ is a real-valued matrix for dimensions $2$ and $4$, which implies that the real and imaginary parts of $\tilde{\mathbf{y}}_{m,k}$ in (\ref{eq:ML_bit_metrics_2}) can be separated, and only the part corresponding to the coded bit is required for calculating one bit metric of the Viterbi decoder. As a result, the decoding complexity of BICMB-PC can be further reduced. Assume that square $M$-QAM is used, whose real and imaginary parts are Gray coded separately as two $\sqrt{M}$-PAM signals. Define $\Re [ \xi_{c_{k'}}^{n,j}]$ and $\Im [ \xi_{c_{k'}}^{n,j}]$ as the signal sets of the real and the imaginary axes of $\xi_{c_{k'}}^{n,j}$, respectively. Therefore, the ML bit metrics in (\ref{eq:ML_bit_metrics_2}) can be further simplified for dimensions $2$ and $4$ as
\begin{align}
\Gamma^{(m,n),j}(\mathbf{Y}_k, c_{k'}) = \min_{\Re[\mathbf{{x}}] \in \Re[\xi_{c_{k'}}^{n,j}]} \| \Re[\tilde{\mathbf{y}}_{m,k}] - \mathbf{R} \Re[\mathbf{x}] \|^2, \label{eq:ML_bit_metrics_real}
\end{align}
if the bit position of $c_{k'}$ is on the real part, or 
\begin{align}
\Gamma^{(m,n),j}(\mathbf{Y}_k, c_{k'}) = \min_{\Im[\mathbf{{x}}] \in \Im[\xi_{c_{k'}}^{n,j}]} \| \Im[\tilde{\mathbf{y}}_{m,k}] - \mathbf{R} \Im[\mathbf{x}] \|^2, \label{eq:ML_bit_metrics_imag}
\end{align}
if the bit position of $c_{k'}$ is on the imaginary part, where $\Re [ \tilde{\mathbf{y}}_{m,k}]$ and $\Im [ \tilde{\mathbf{y}}_{m,k}]$ denote the real and imaginary parts of $\tilde{\mathbf{y}}_{m,k}$, respectively. For (\ref{eq:ML_bit_metrics_real}) and (\ref{eq:ML_bit_metrics_imag}), the worst-case decoding complexity is only $\mathcal{O}(M^{\frac{D}{2}}-0.5)$ when SD with rounding (or quantization) procedure for the last layer is employed, which is much lower than BICMB-FP of $\mathcal{O}(M^{D}-1)$. 

Note that in \cite{Li_GCMB}, for the uncoded SVD systems with both full diversity and full multiplexing, the decoding problem of PCMB is to solve two $D$-dimensional real-valued problems, while that of FPMB is an $D$-dimensional complex-valued problem in the case of dimensions $2$ and $4$. In the convolutional coded case as shown in this paper, the metric calculation problem of BICMB-PC is only one $D$-dimensional real-valued problem compared to one $D$-dimensional complex-valued problem of BICMB-FP for dimensions $2$ and $4$. Therefore, the complexity reduction from BICMB-FP to BICMB-PC is greater than the reduction from FPMB to PCMB achieved in \cite{Li_GCMB} for dimensions $2$ and $4$. Moreover, since the precoded part of BICMB-PP could be considered as a smaller dimensional BICMB-FP, BICMB-PC of dimensions $2$ and $4$ could be applied to replace the precoded part and reduce the complexity for BICMB-PP. 

Note that (\ref{eq:ML_bit_metrics_real}) and (\ref{eq:ML_bit_metrics_imag}) could be further simplified with the trade-off of some performance degradation. One way is to replace the square of the difference with the absolute value. Another way is to multiply $\mathbf{R}^{-1}$ with $\Re[\tilde{\mathbf{y}}_{m,k}]$ and $\Im[\tilde{\mathbf{y}}_{m,k}]$, which is actually a modified Zero-Forcing (ZF) method. Since we focus on ML decoding, further details of the trade-off between performance and complexity are not provided.

%% file: Results.tex
\section{Simulation Results} \label{sec:Results}




Fig. \ref{fig:ber_snr_2x2} and Fig. \ref{fig:ber_snr_4x4} show BER-SNR performance comparison of BICMB-PC and BICMB-FP using different modulation schemes for $R_c=2/3$, $2\times2$ systems, and $R_c=4/5$, $4\times4$ systems, respectively. The constellation precoders for FPMB are selected as the best ones introduced in \cite{Park_CPB}. Simulation results show that BICMB-PC and BICMB-FP achieve almost the same performance for both dimensions with all the considered modulation schemes. Moreover, the worst-case decoding complexity of $\mathcal{O}(M^{\frac{D}{2}}-0.5)$ to get one bit metric for BICMB-PC is much lower than $\mathcal{O}(M^D-1)$ for BICMB-FP. 

In order to measure the decoding complexity, the average number of real multiplications which are the most expensive operations in terms of machine cycles, for acquiring one bit metric is calculated at different SNR for BICMB-PC and BICMB-FP, respectively.
In \cite{Li_RCSD_J}, \cite{Li_RC_BICMB_CP}, an efficient reduced complexity decoding technique was introduced for BICMB-FP, which is applied in this paper. For fair comparisons, a similar decoding technique is applied to BICMB-PC. Fig. \ref{fig:complexity_coded} shows the complexity comparisons for BICMB-PC and BICMB-FP in dimensions of $2$ and $4$ using $64$-QAM. For the dimension of $2$, the complexity of BICMB-PC is $0.8$ and $0.5$ orders of magnitude lower than BICMB-FP at low and high SNR respectively. In the dimension of $4$ case, the improvements reach $2.2$ and $1.6$ orders of magnitude.

Additionally, Fig. \ref{fig:complexity_uncoded} shows the complexity comparisons for uncoded full-diversity full-multiplexing SVD systems of FPMB and PCMB in dimensions of $2$ and $4$ using $64$-QAM. In \cite{Li_RCSD_J}, \cite{Li_RCSD}, an efficient reduced complexity SD technique was introduced, which is applied in this paper. For the dimension of $2$, the complexity of GCMB is $0.5$ order of magnitude lower than FPMB at low SNR and similar to FPMB at high SNR. In the dimension of $4$ case, the complexity of GCMB is $1.7$ orders of magnitude lower than FPMB at low SNR and similar to FPMB at high SNR. Fig. \ref{fig:complexity_coded} and Fig. \ref{fig:complexity_uncoded} show that the complexity reduction from BICMB-FP to BICMB-PC is greater than the reduction from FPMB to PCMB achieved in \cite{Li_GCMB} for dimensions $2$ and $4$.

\ifCLASSOPTIONonecolumn
\begin{figure}[!m]
\centering
\scalebox{.7}{\includegraphics{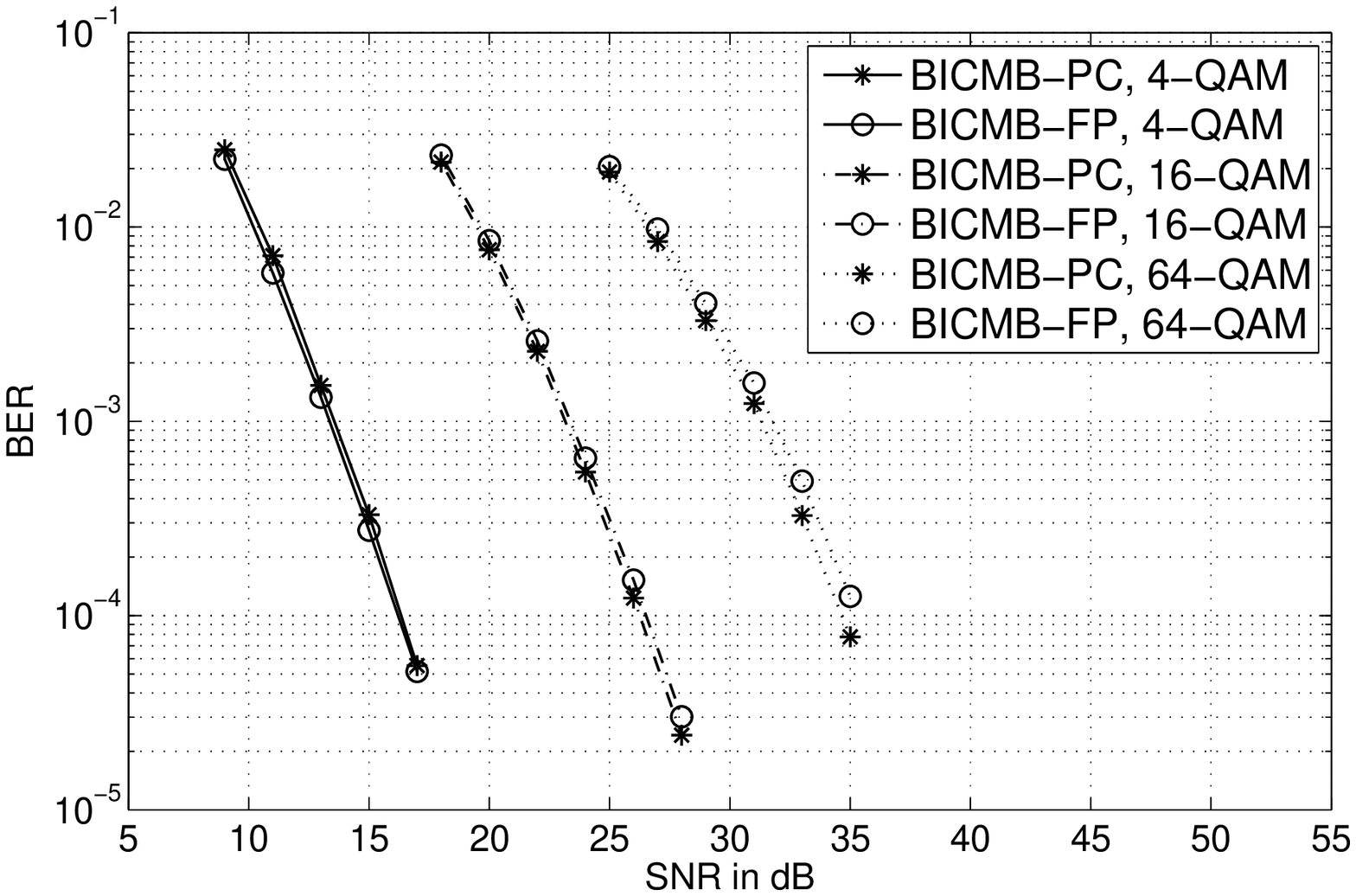}}
\caption{BER vs. SNR for BICMB-PC and BICMB-FP for $R_c=2/3$, $2\times2$ systems.}
\label{fig:ber_snr_2x2}
\end{figure}

\begin{figure}[!m]
\centering
\scalebox{.7}{\includegraphics{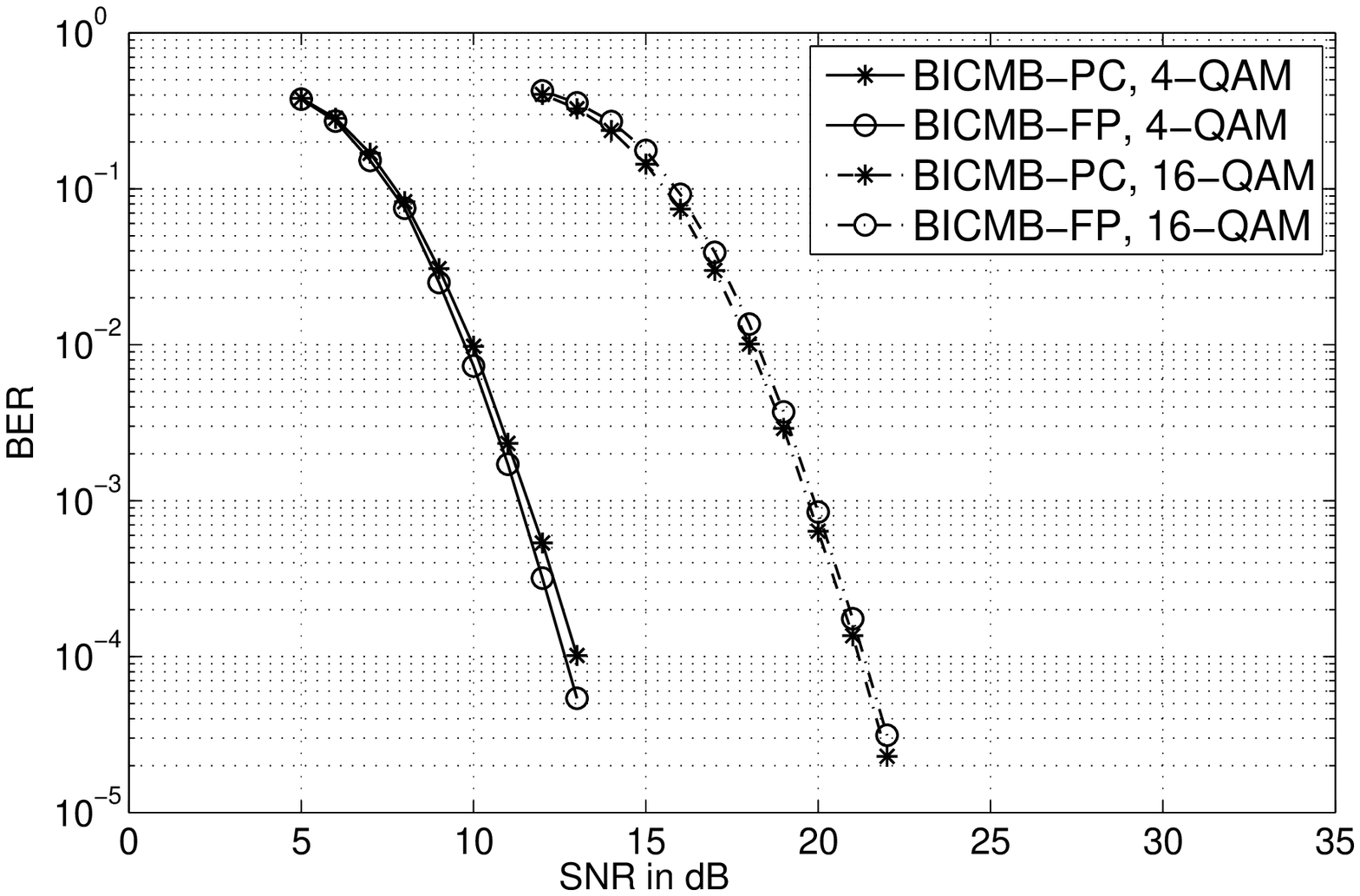}}
\caption{BER vs. SNR for BICMB-PC and BICMB-FP for $R_c=4/5$, $4\times4$ systems.}
\label{fig:ber_snr_4x4}
\end{figure}

\begin{figure}[!m]
\centering
\scalebox{.7}{\includegraphics{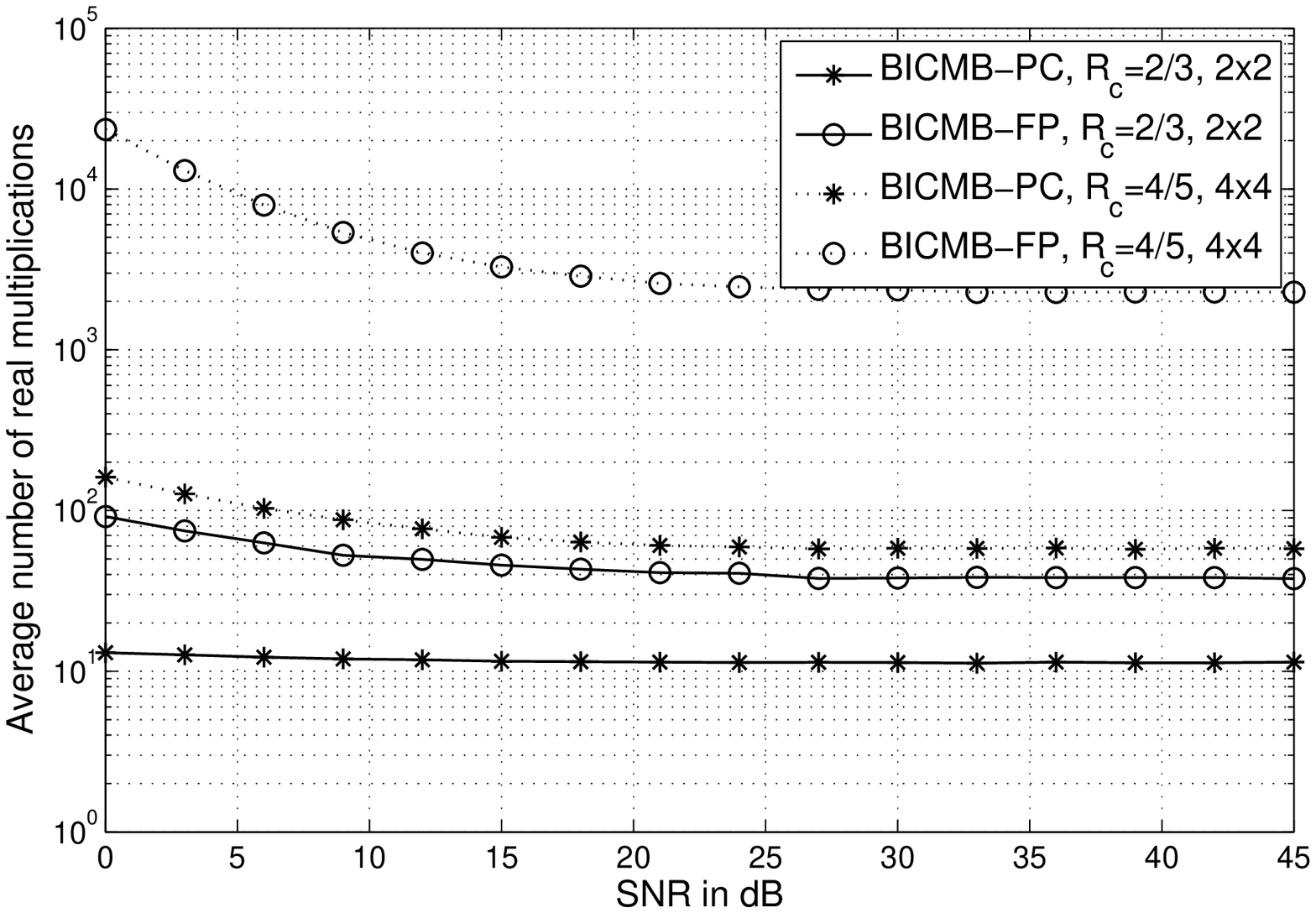}}
\caption{Average number of real multiplications vs. SNR for BICMB-PC and BICMB-FP using $64$-QAM.}
\label{fig:complexity_coded}
\end{figure}

\begin{figure}[!m]
\centering
\scalebox{.7}{\includegraphics{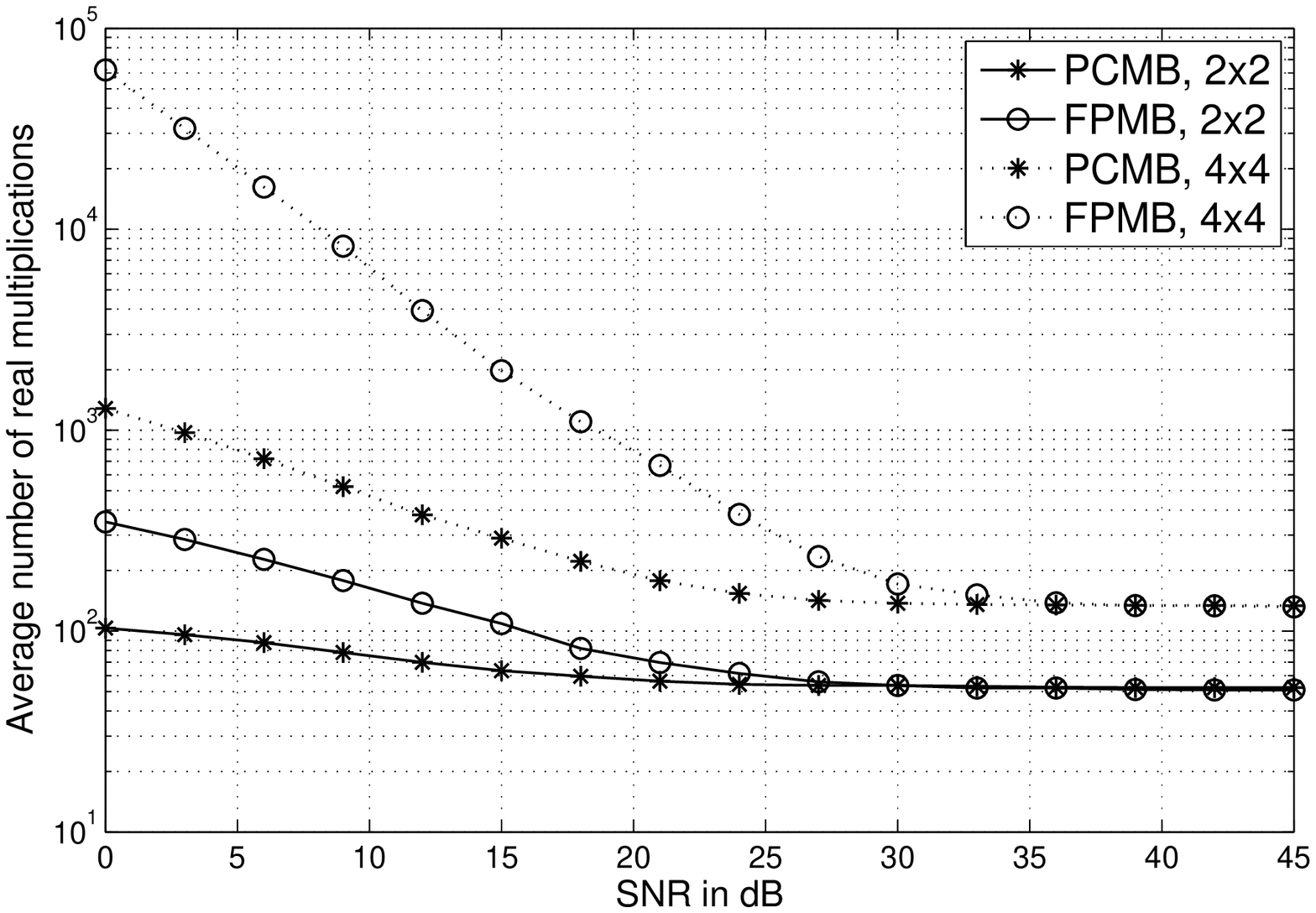}}
\caption{Average number of real multiplications vs. SNR for PCMB and FPMB using $64$-QAM.}
\label{fig:complexity_uncoded}
\end{figure}

\else
\begin{figure}[!t]
\centering
\scalebox{.4}{\includegraphics{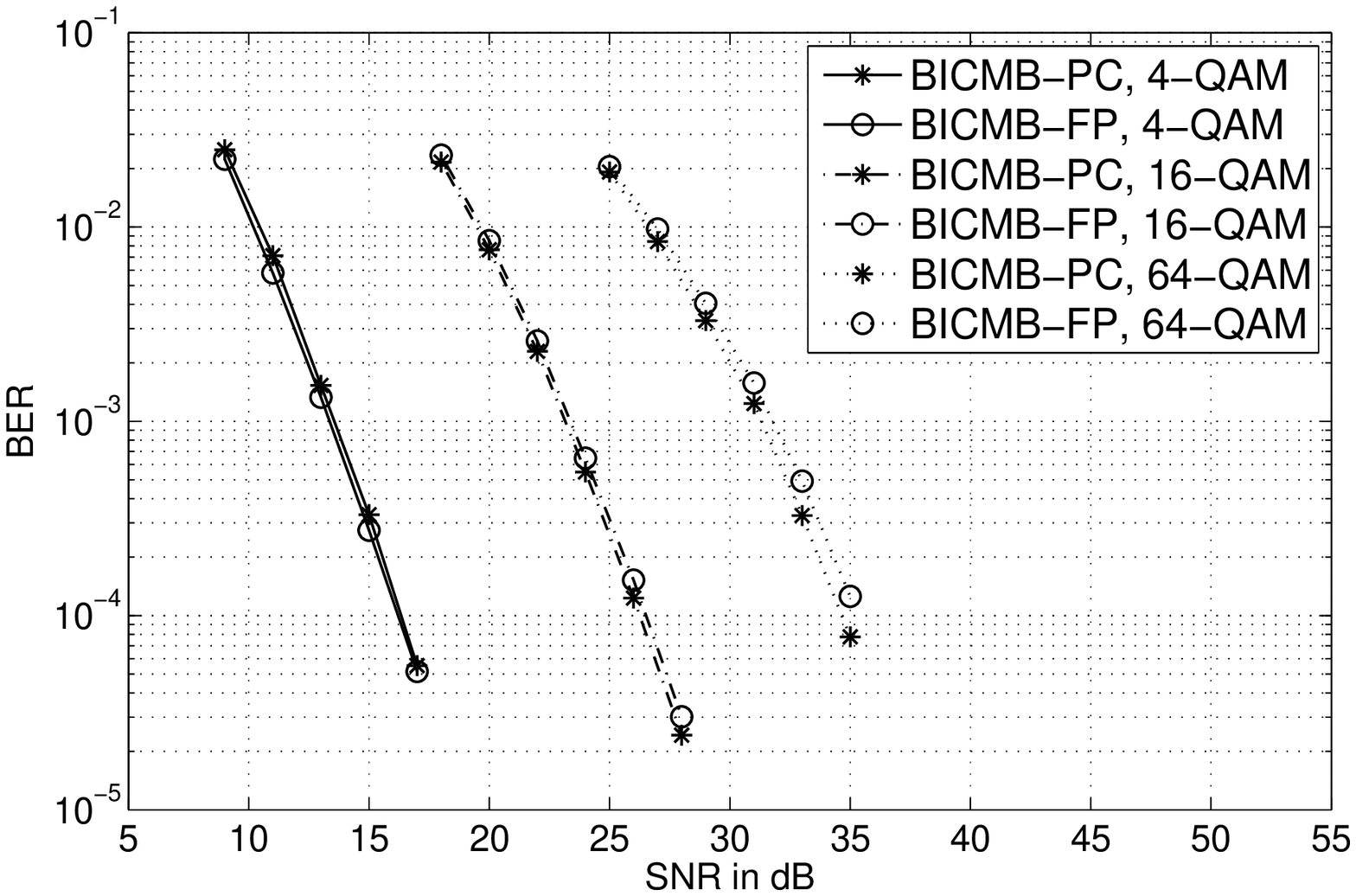}}
\caption{BER vs. SNR for BICMB-PC and BICMB-FP for $R_c=2/3$, $2\times2$ systems.}
\label{fig:ber_snr_2x2}
\end{figure}

\begin{figure}[!t]
\centering
\scalebox{.4}{\includegraphics{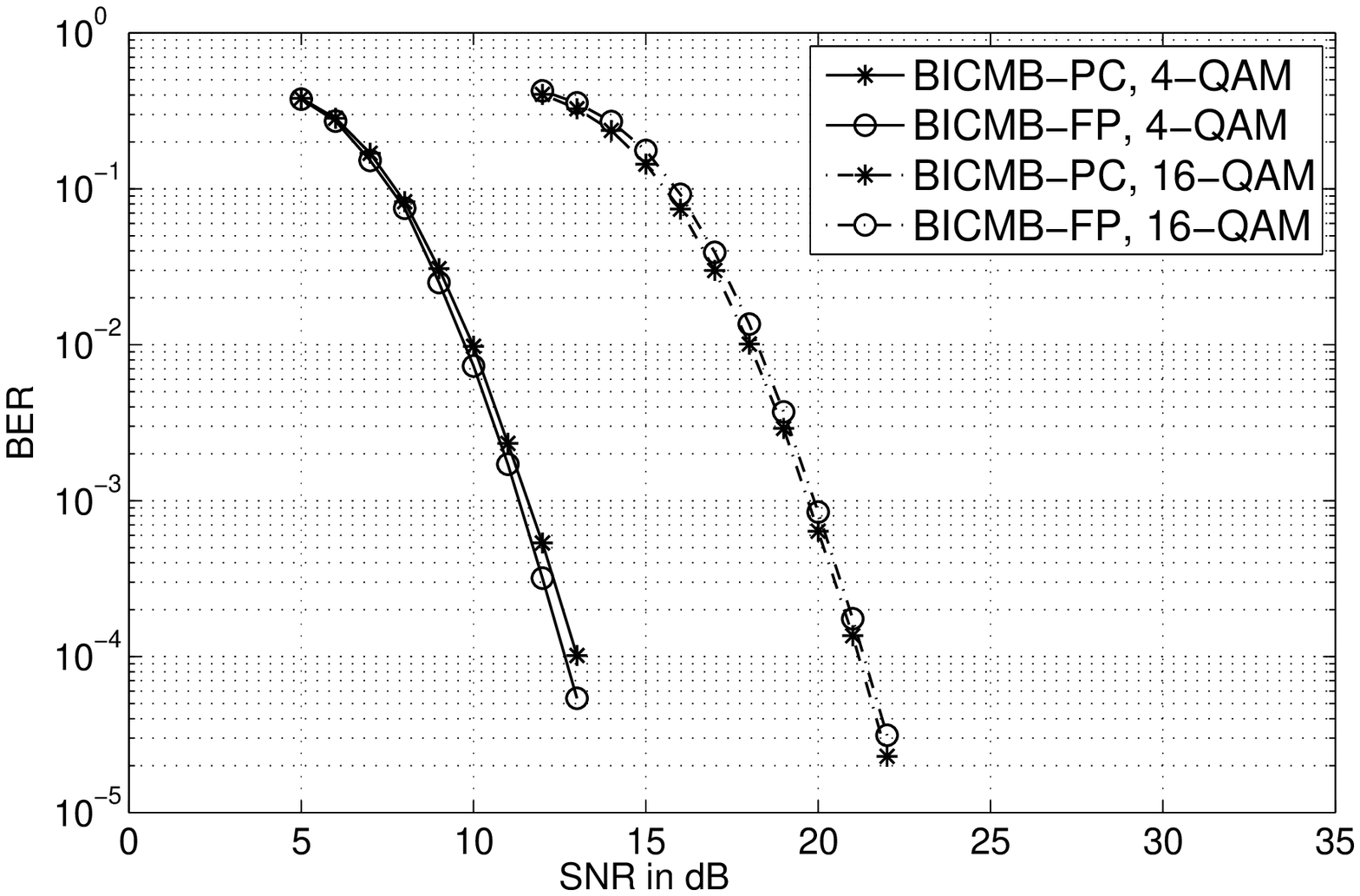}}
\caption{BER vs. SNR for BICMB-PC and BICMB-FP for $R_c=4/5$, $4\times4$ systems.}
\label{fig:ber_snr_4x4}
\end{figure}

\begin{figure}[!t]
\centering
\scalebox{.4}{\includegraphics{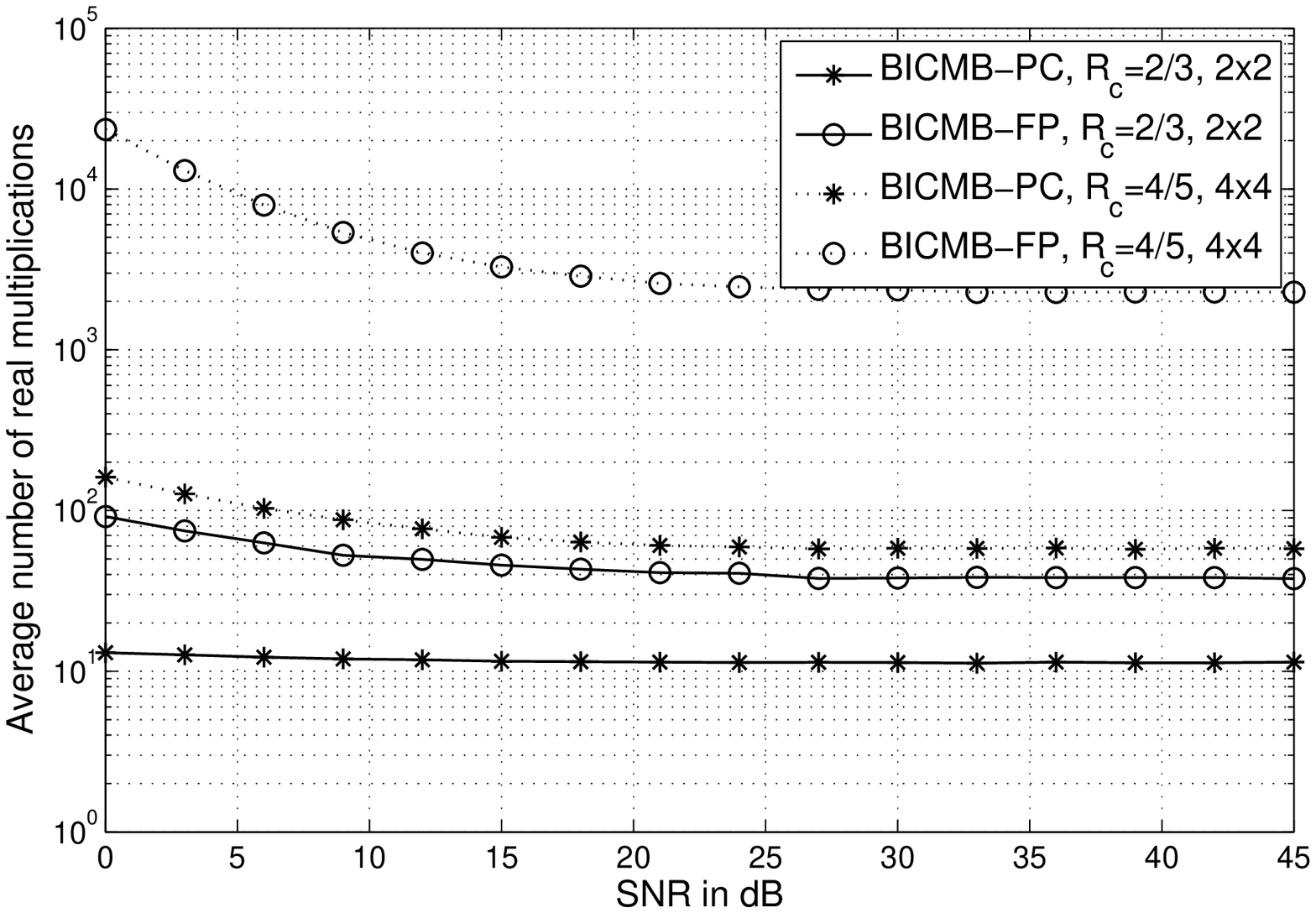}}
\caption{Average number of real multiplications vs. SNR for BICMB-PC and BICMB-FP using $64$-QAM.}
\label{fig:complexity_coded}
\end{figure}

\begin{figure}[!t]
\centering
\scalebox{.4}{\includegraphics{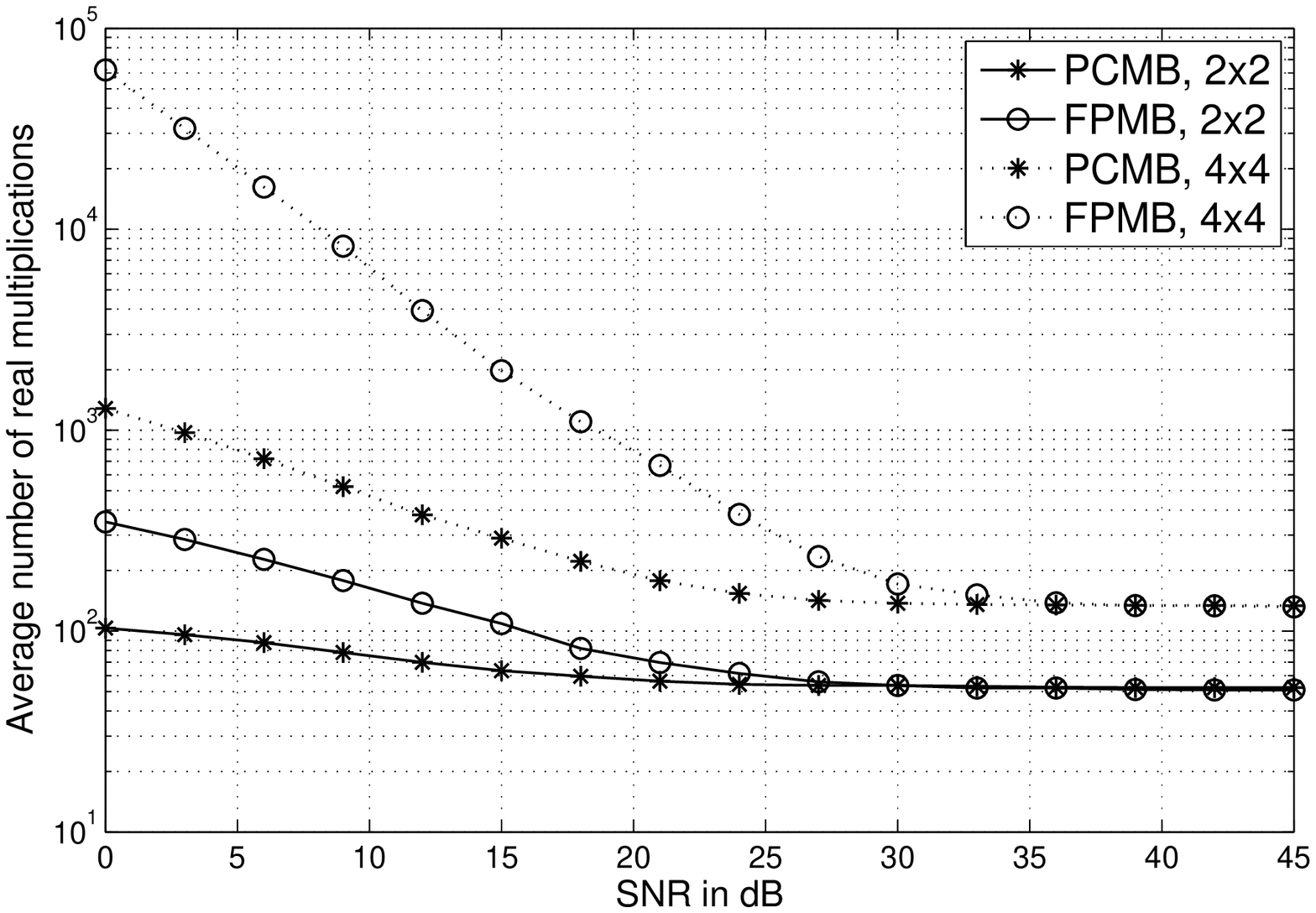}}
\caption{Average number of real multiplications vs. SNR for PCMB and FPMB using $64$-QAM.}
\label{fig:complexity_uncoded}
\end{figure}

\fi

%% file: Conclusion.tex
\section{Conclusion} \label{sec:Conclusion}

In this paper, BICMB-PC which combines PSTBC and multiple beamforming technique is proposed. It is a new technique with both full diversity and full multiplexing for convolutional coded SVD-MIMO systems. Diversity analysis and decoding complexity analysis are provided. It is shown that BICMB-PC achieves a similar BER performance to BICMB-FP, which is also a full-diversity full-multiplexing technique for convolutional coded SVD-MIMO systems. Particularly, for dimensions $2$ and $4$, because only one of the real or imaginary part of the received signal is required to calculate one bit metric for the Viterbi decoder, the worst-case decoding complexity of $\mathcal{O}(M^{\frac{D}{2}}-0.5)$ for BICMB-PC is much lower than $\mathcal{O}(M^D-1)$ for BICMB-FP, which provides the advantage of BICMB-PC. Compared to the uncoded system, the complexity reduction from BICMB-FP to BICMB-PC is greater than the reduction from FPMB to PCMB achieved in \cite{Li_GCMB} for dimensions $2$ and $4$. Furthermore, since the precoded part of BICMB-PP could be considered as a smaller dimensional BICMB-FP, BICMB-PC of dimensions $2$ and $4$ could be applied to replace the precoded part and reduce the complexity for BICMB-PP. 

%% file: Main.bbl
\begin{thebibliography}{10}
\providecommand{\url}[1]{#1}
\csname url@samestyle\endcsname
\providecommand{\newblock}{\relax}
\providecommand{\bibinfo}[2]{#2}
\providecommand{\BIBentrySTDinterwordspacing}{\spaceskip=0pt\relax}
\providecommand{\BIBentryALTinterwordstretchfactor}{4}
\providecommand{\BIBentryALTinterwordspacing}{\spaceskip=\fontdimen2\font plus
\BIBentryALTinterwordstretchfactor\fontdimen3\font minus
  \fontdimen4\font\relax}
\providecommand{\BIBforeignlanguage}[2]{{%
\expandafter\ifx\csname l@#1\endcsname\relax
\typeout{** WARNING: IEEEtran.bst: No hyphenation pattern has been}%
\typeout{** loaded for the language `#1'. Using the pattern for}%
\typeout{** the default language instead.}%
\else
\language=\csname l@#1\endcsname
\fi
#2}}
\providecommand{\BIBdecl}{\relax}
\BIBdecl

\bibitem{Paulraj_ST}
A.~Paulraj, R.~Nabar, and D.~Gore, \emph{Introduction to Space-Time Wireless
  Communication}.\hskip 1em plus 0.5em minus 0.4em\relax Cambridge University
  Press, 2003.

\bibitem{Zheng_DM}
L.~Zheng. and D.~Tse, ``{Diversity and Multiplexing: a Fundamental Tradeoff In
  Multiple-antenna Channels},'' \emph{{IEEE} Trans. Inf. Theory}, vol.~49,
  no.~5, pp. 1073--1096, May 2003.

\bibitem{Jafarkhani_STC}
H.~Jafarkhani, \emph{Space-Time Coding: Theory and Practice}.\hskip 1em plus
  0.5em minus 0.4em\relax Cambridge University Press, 2005.

\bibitem{Sengul_DA_SMB}
E.~Sengul, E.~Akay, and E.~Ayanoglu, ``{Diversity Analysis of Single and
  Multiple Beamforming},'' \emph{{IEEE} Trans. Commun.}, vol.~54, no.~6, pp.
  990--993, Jun. 2006.

\bibitem{Akay_BICMB}
E.~Akay, E.~Sengul, and E.~Ayanoglu, ``{Bit-Interleaved Coded Multiple
  Beamforming},'' \emph{{IEEE} Trans. Commun.}, vol.~55, no.~9, pp. 1802--1811,
  Sep. 2007.

\bibitem{Akay_On_BICMB}
\BIBentryALTinterwordspacing
E.~Akay, H.~J. Park, and E.~Ayanoglu. (2008) {On "Bit-Interleaved Coded
  Multiple Beamforming"}. arXiv: 0807.2464. [Online]. Available:
  \url{http://arxiv.org}
\BIBentrySTDinterwordspacing

\bibitem{Park_DA_BICMB}
H.~J. Park and E.~Ayanoglu, ``{Diversity Analysis of Bit-Interleaved Coded
  Multiple Beamforming},'' in \emph{Proc. IEEE ICC 2009}, Dresden, Germany,
  Jun. 2009.

\bibitem{Park_DA_BICMB_J}
------, ``{Diversity Analysis of Bit-Interleaved Coded Multiple Beamforming},''
  \emph{{IEEE} Trans. Commun.}, vol.~58, no.~8, pp. 2457--2463, Aug. 2010.

\bibitem{Mohammed_DIA}
S.~K. Mohammed, E.~Viterbo, Y.~Hong, and A.~Chockalingam, ``{Precoding by
  Pairing Subchannels to Increase MIMO Capacity with Discrete Input
  Alphabets},'' \emph{{IEEE} Trans. Inf. Theory}, vol.~57, no.~7, pp.
  4156--4169, Jul. 2011.

\bibitem{Mohammed_XY}
------, ``{MIMO Precoding with X- and Y-Codes},'' \emph{{IEEE} Trans. Inf.
  Theory}, vol.~57, no.~6, pp. 3542--3566, Jun. 2011.

\bibitem{Park_CPB}
H.~J. Park and E.~Ayanoglu, ``{Constellation Precoded Beamforming},'' in
  \emph{Proc. IEEE GLOBECOM 2009}, Honolulu, HI, USA, Nov. 2009.

\bibitem{Park_BICMB_CP}
------, ``{Bit-Interleaved Coded Multiple Beamforming with Constellation
  Precoding},'' in \emph{Proc. IEEE ICC 2010}, Cape Town, South Africa, May
  2010.

\bibitem{Park_MB_CP}
H.~J. Park, B.~Li, and E.~Ayanoglu, ``{Multiple Beamforming with Constellation
  Precoding: Diversity Analysis and Sphere Decoding},'' in \emph{Proc. IEEE ITA
  2010}, San Diego, CA, USA, Apr. 2010.

\bibitem{Park_CPMB}
------, ``{Constellation Precoded Multiple Beamforming},'' \emph{{IEEE} Trans.
  Commun.}, vol.~59, no.~5, pp. 1275--1286, May 2011.

\bibitem{Srinivas_COISM_CSI}
K.~Srinivas, R.~Koilpillai, S.~Bhashyam, and K.~Giridhar, ``{Co-Ordinate
  Interleaved Spatial Multiplexing with Channel State Information},''
  \emph{{IEEE} Trans. Wireless Commun.}, vol.~6, no.~8, pp. 2755--2762, Jun.
  2009.

\bibitem{Oggier_PSTBC}
F.~Oggier, G.~G.~Rekaya, J.-C. Belfiore, and E.~Viterbo, ``{Perfect Space-Time
  Block Codes},'' \emph{{IEEE} Trans. Inf. Theory}, vol.~52, no.~9, pp.
  3885--3902, 2006.

\bibitem{Elia_PSTBC}
P.~Elia, B.~A. Sethuraman, and P.~V. Kumar, ``{Perfect Space-Time Codes with
  Minimum and Non-Minimum Delay for any Number of Antennas},'' in \emph{Proc.
  WIRELESSCOM 2005}, vol.~1, Sheraton Maui Resort, HI, USA, Jun. 2005, pp.
  722--727.

\bibitem{Berhuy_PSTC}
G.~Berhuy and F.~Oggier, ``{On the Existence of Perfect Space-Time Codes},''
  \emph{{IEEE} Trans. Inf. Theory}, vol.~55, no.~5, pp. 2078--2082, May 2009.

\bibitem{Belfiore_GC}
J.-C. Belfiore, G.~Rekaya, and E.~Viterbo, ``{The Golden Code: A $2\times2$
  Full-Rate Space-Time Code With Nonvanishing Determinants},'' \emph{{IEEE}
  Trans. Inf. Theory}, vol.~51, pp. 1432--1436, Apr. 2005.

\bibitem{Dayal_STC}
P.~Dayal and M.~K. Varanasi, ``{An Optimal Two Transmit Antenna Space-Time Code
  and Its Stacked Extensions},'' \emph{{IEEE} Trans. Inf. Theory}, vol.~51,
  no.~12, pp. 4348--4355, Dec. 2005.

\bibitem{IEEE_802_16e}
``{IEEE 802.16e-2005: IEEE Standard for Local and Metropolitan Area Network -
  Part 16: Air Interface for Fixed and Mobile Broadband Wireless Access Systems
  - Amendment 2: Physical Layer and Medium Access Control Layers for Combined
  Fixed and Mobile Operation in Licensed Bands},'' Feb. 2006.

\bibitem{Li_GCMB}
B.~Li and E.~Ayanoglu, ``{Golden Coded Multiple Beamfoming},'' in \emph{Proc.
  IEEE GLOBECOM 2010}, Miami, FL, USA, Dec. 2010.

\bibitem{Haccoun_PCC}
D.~Haccoun and G.~Begin, ``{High-Rate Punctured Convolutional Codes for Viterbi
  and Sequential Decoding},'' \emph{{IEEE} Trans. Commun.}, vol.~37, no.~11,
  pp. 1113--1125, Nov. 1989.

\bibitem{Forney_EM}
G.~J. Forney, R.~Gallager, G.~Lang, F.~Longstaff, and S.~Qureshi, ``{Efficient
  Modulation for Band-Limited Channels},'' \emph{{IEEE} J. Sel. Areas Commun.},
  vol.~2, no.~5, pp. 632--647, Sep. 1984.

\bibitem{Lin_ECC}
S.~Lin and D.~J. Costello, \emph{Error Control Coding: Fundamentals and
  Applications}.\hskip 1em plus 0.5em minus 0.4em\relax Prentice Hall, Second
  Edition, 2004.

\bibitem{Caire_BICM}
G.~Caire, G.~Taricco, and E.~Biglieri, ``{Bit-Interleaved Coded Modulation},''
  \emph{{IEEE} Trans. Inf. Theory}, vol.~44, no.~3, pp. 927--946, May 1998.

\bibitem{Park_UP_MPDF}
H.~J. Park and E.~Ayanoglu, ``{An Upper Bound to the Marginal PDF of the
  Ordered Eigenvalues of Wishart Matrices and Its Application to MIMO Diversity
  Analysis},'' in \emph{Proc. IEEE ICC 2010}, Cape Town, South Africa, May
  2010.

\bibitem{Jalden_SD}
J.~Jald{\'{e}}n and B.~Ottersten, ``{On the Complexity of Sphere Decoding in
  Digital Communications},'' \emph{{IEEE} Trans. Signal Process.}, vol.~53,
  no.~4, pp. 1474--1484, Apr. 2005.

\bibitem{Sinnokrot_STBC_LMLDC}
M.~O. Sinnokrot, ``Space-time block codes with low maximum likelihood decoding
  complexity,'' Ph.D. dissertation, Georgia Institute of Technology, Dec. 2009.

\bibitem{Li_RCSD_J}
B.~Li and E.~Ayanoglu, ``{Reduced Complexity Sphere Decoding},'' \emph{Wiley
  Wireless Communications and Mobile Computing}, vol.~11, no.~12, pp.
  1518--1527, Dec. 2011.

\bibitem{Li_RC_BICMB_CP}
------, ``{Reduced Complexity Decoding for Bit-Interleaved Coded Multiple
  Beamforming with Constellation Precoding},'' in \emph{Proc. IEEE IWCMC 2011},
  Istanbul, Turkey, Jul. 2011.

\bibitem{Li_RCSD}
------, ``{Reduced Complexity Sphere Decoding},'' in \emph{Proc. IEEE IWCMC
  2011}, Istanbul, Turkey, Jul. 2011.

\end{thebibliography}
